\documentclass[twocolumn,superscriptaddress,amsmath,amssymb,showpacs,prb]{revtex4-1}

\usepackage{graphicx}
\usepackage{color}
\renewcommand{\phi}{\varphi}

\begin{document}

\title{LDA+$\textit{U}_\text{sc}$ calculations  of phase relations in FeO}

\author{Yang Sun}
    \affiliation{Department of Applied Physics and Applied Mathematics, Columbia University, New York, NY, 10027, USA}
\author{Matteo Cococcioni}
	\affiliation{Department of physics, University of Pavia, 27100 Pavia, Italy}
\author{Renata M. Wentzcovith}
	\affiliation{Department of Applied Physics and Applied Mathematics, Columbia University, New York, NY, 10027, USA}
	\affiliation{Department of Earth and Environmental Sciences, Columbia University, New York, NY, 10027, USA}
	\affiliation{Lamont–Doherty Earth Observatory, Columbia University, Palisades, NY, 10964, USA}

\date{May 23, 2020}

\begin{abstract}

Using the LDA+$\textit{U}_\text{sc}$ method, we present calculations phase relations of iron monoxides involving five polytypes in multiple spin-state configurations. The Hubbard parameter $U$ is determined self-consistently simultaneously with the occupation matrix and structures at arbitrary pressures. The Hubbard parameter strongly depends on pressure, structure, and spin state. Comparison with experimental structural data indicates the LDA+$\textit{U}_\text{sc}$ can predict structure, compression curves, phase relations, and transition pressures very well for the insulating B1 and iB8 states. However, it requires additional calculations using the Mermin functional that includes the electronic entropic contribution to the free energy to obtain an nB8 metallic state and a consistent iB8 to nB8 insulator-to-metal transition pressure.

\end{abstract}

\maketitle

\section{Introduction}

Fe-O is one of the most fundamental components of the Earth’s mantle and possibly also of the Earth’s core. Phase relations in FeO are essential to determining the state of iron at the extreme pressures of deep planetary environments. Several challenging experiments under extreme conditions have been performed in FeO in the last decades \cite{1,2,3,4,5,6}. FeO is also an archetypical strongly correlated material \cite{7}. From the materials simulations viewpoint, it is an important test case in the development of methods to predict phase stability at arbitrary conditions.  

Modern computational materials discovery methods have shown to be an ideal tool to discover new materials at extreme conditions \cite{8,9}. However, these methods have achieved only limited success for the Fe-O system \cite{10,11,12}. With few exceptions \cite{6}, they mostly predicted phases that are not observed experimentally. The presence of localized and strongly correlated $3d$ electrons in iron prevents successful applications of $ab initio$ methods using conventional exchange-correlation functionals such as the local density approximation (LDA) or the generalized gradient approximation (GGA) to the Fe-O system. For example, it is a well-known fact that conventional band-structure calculations with the LDA or GGA incorrectly give a metallic ground state for FeO in the B1 structure. This problem has been solved by using many-body electronic structure methods that address the strongly correlated state of the 3d electrons \cite{13,14,15,16,17}. For example, by including the Hubbard correction in these standard DFT calculations, DFT+$\textit{U}$ calculations open the Hubbard gap and produce the insulating state of B1-type FeO \cite{13,18}. The reliability of the DFT+$\textit{U}$ results depends on the Hubbard parameter $\textit{U}$. While it has been argued that $\textit{U}$ should be determined by first-principles \cite{14}, self-consistently \cite{19}, and be structure and spin-state dependent \cite{20,21,22,23,24}, many studies have employed a constant and semi-arbitrary $\textit{U}$ value or tuned $\textit{U}$ to match experimental observations of some sort. While useful in providing insights into the electronic structure problem (see, e.g., \cite{25,26}), these calculations have not fully explored the predictive power of this method. The full dependence of $\textit{U}$ on pressure/volume, structure, spin state, chemistry, etc., must be computed if one is to make predictions at extreme conditions in the presence of dissociation or recombination reactions. On the other hand, determining the Hubbard parameter is a non-trivial task \cite{14,19}. Still, a recent implementation based on density-functional perturbation theory (DFPT) and monochromatic perturbations \cite{27} dramatically facilitates the calculation of $U$ by decreasing the computational effort by orders of magnitudes.

In this work, we focus on the iron monoxide, which is the most frequently studied of the Fe-O compounds. We consider iron monoxide in five polytypes and investigate the predictive power of self-consistent $\textit{U}_\text{sc}$ calculations to reproduce its experimental phase relations. We compute relative phase stabilities, equations of state, and transition pressures, and compare them with experimental information. 

In the next session, we review previous high-pressure works on iron monoxide. Section III discusses the methods used to compute the Hubbard $\textit{U}$ and total energies while section IV presents and discusses our results. Conclusions are presented in Section V.

\section{Review of previous works}

FeO has a complex phase diagram. At ambient conditions, FeO forms a cubic NaCl-type B1 structure. At higher pressure or lower temperature, the symmetry reduces to rhombohedra (rB1) with antiferromagnetic ordering along [111] \cite{1}. The phase boundary between the B1 and rB1 was measured at and above room temperature \cite{2,28,29}. While the rB1 phase is usually considered to be the ground state at low temperatures up to $\sim$ 100 GPa \cite{30}, Fjellvag $\textit{et al.}$ \cite{31} observed further symmetry reduction from rhombohedral to monoclinic at $T=10\ K$. This lower symmetry also features a local Jahn-Teller distortion of FeO$_\text{6}$ octahedron that produces four short and two long bonds \cite{31}. With DFT+$\textit{U}$ calculations, Cococcioni and de Gironcoli \cite{14} described the reduced symmetry of the rhombohedral lattice by moving the minority-spin $\textit{d}$ electron from the a$_{\text{1g}}$ (z$^2$) orbital to one of the degenerated t$_{\text{2g}}$ orbitals. Gramsch $\textit{et al.}$ \cite{25} explained the monoclinic symmetry could lead to the splitting of the e$_{\text{g}}$ pair into a$_{\text{g}}$ and b$_{\text{g}}$ states and the a$_{\text{g}}$ orbital occupancy could be stabilized by the Hubbard parameter. However, phase stability and phase transitions in the monoclinic phase were not addressed in these studies.

\begin{figure}[t]
\includegraphics[width=0.45\textwidth]{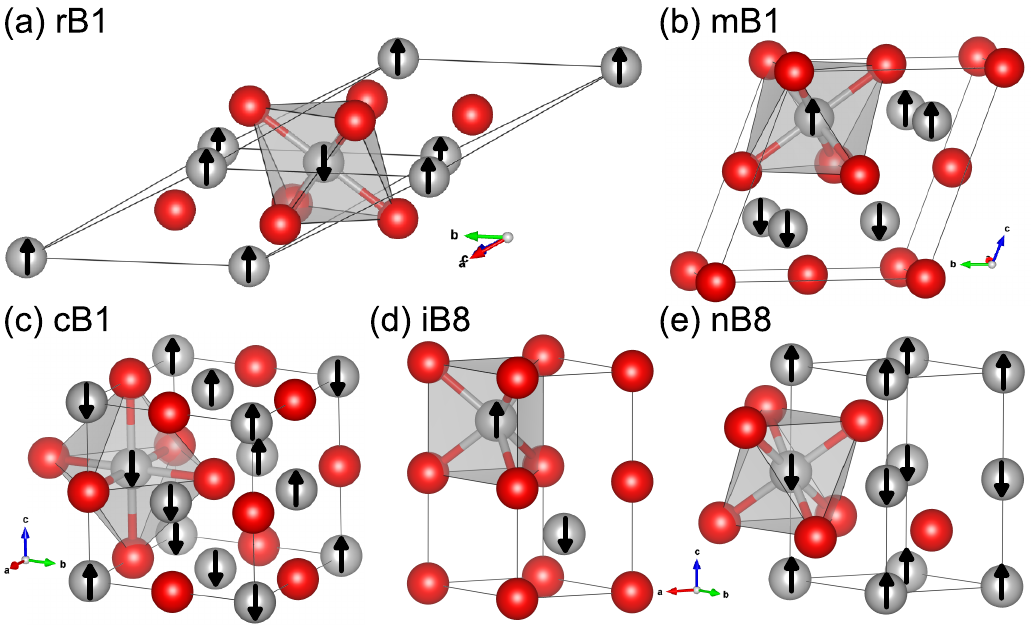}
\caption{\label{fig:fig1} Crystal structures and magnetic configurations for (a) rhombohedral B1, (b) monoclinic B1, (c) cubic B1phases, (d) inverse B8 and (e) normal B8. Iron ions are shown in grey and oxygen ions in red. Arrows indicate the spin direction of various iron ions. The local iron coordination is indicated by the surrounding polyhedron.}
\end{figure}

Fei and Mao found experimentally that under pressure, the rB1 phase transforms to the B8 phase above 90 GPa at 600 K \cite{2}. The equilibrium phase boundary between rB1 and B8 was found near 105 GPa at room temperature\cite{30,32}. Two polytypes of B8 FeO have been reported \cite{28,33,34}: a) the normal NiAs-type B8 structure (nB8 hereafter) and b) the inverse B8 structure (iB8 hereafter). The main structural difference between nB8 and iB8 phases is that the Fe and O positions are swapped. These structures have different Fe coordination polyhedra. In the iB8 structure, the coordination polyhedron is a triangular prism, while in the nB8 structure is an octahedron (see Fig.~\ref{fig:fig1}). A high-spin (HS) to low-spin (LS) state change was also found in FeO at high-pressures. This transition pressure has been debated for a long time. Pasternak $et\ al.$ suggested that the HS state transforms into the LS state at 90 GPa \cite{35}. Later, Badro $\textit{et al.}$ \cite{36} reported the HS could be preserved up to 143 GPa at room temperature. Mattila $et\ al.$ \cite{37} found that LS FeO existed above 140 GPa. Most recently, Ohta $et\ al.$ demonstrated that the structural transition from iB8 to nB8, as well as the HS-LS state change and the insulator-metal transition, all happened concurrently near 120 GPa with little temperature dependence \cite{38,39}. 

Theoretical calculations by Sherman and Jansen \cite{40} reported a structural transition from rB1 to B8 at 130 GPa using the LDA. Cohen $et\ al.$ \cite{3} found the magnetic collapse in the B1 structure at 100 GPa with LDA or at 200 GPa with GGA. Mazin \cite{33} indicated that the GGA and LDA exchange-correlation functionals could not describe the relative stability between iB8 and nB8. Persson $et\ al.$ \cite{26} obtained the spin state change in B1 near 200 GPa using GGA+U with constant $\textit{U}$ (3eV and 5eV). The metal-insulator transitions (MIT) of FeO was mainly studied by the combination of DFT and dynamical mean-field theory (DFT+DMFT). Shorikov $et\ al.$ \cite{16} predicted the MIT for B1-HS at 60GPa at room temperature. The prediction does not agree with experiments. Later, Ohta $et\ al.$ performed DFT+DMFT calculations for the B1 phase and obtained consistent results with experiments that the B1 phase metallization happened at high temperature and high pressure, and it was related to the HS-LS transition \cite{29}. Leonov \cite{41} also showed that DFT+DMFT could describe the MIT for B1, and the transition pressure quantitatively changes with the Hubbard parameter.

\section{Methods}

Here, LDA+$\textit{U}$ calculations were performed using the simplified formulation of Dudarev $\textit{et al.}$ \cite{42} as implemented in the Quantum ESPRESSO code [43,44]. The LDA was used for the exchange-correlation functional. We have used ultra-soft pseudopotentials \cite{45} for Fe and O with valence electronic configurations $3s^23p^63d^{6.5}4s^14p^0$ and $2s^22p^4$ for Fe and O, respectively. Such potentials were generated, tested, and previously used, e.g., in \cite{46}. A kinetic-energy cutoff of 50 Ry for wave functions and 500 Ry for spin-charge density and potentials were used. In all cases, the atomic orbitals were used to construct occupation matrices and projectors in the LDA+$\textit{U}$ scheme.

Three B1 structures and two B8 structures were considered in the present calculations, that is rhombohedra B1 (rB1, space group R$\bar{\text{3}}$m), monoclinic B1 (mB1, space group  C2/m), cubic B1 (cB1, space group Fm$\bar{\text{3}}$m), inverse B8 (iB8, space group P6$_\text{3}$/mmc) and normal B8 (nB8). The structures are plotted in Fig.~\ref{fig:fig1}. A $6\times6\times6$ \textbf{k}-point mesh was used for Brillouin zone integration for all structures, which was sufficient to achieve a convergence of 1 meV/atom in the total energy. The convergence thresholds are 0.01 eV/$\text{\AA}$ for the atomic force, 0.5 kbar for the pressure and $1\times{10}^{-5}$ eV for the total energy.

The Hubbard correction \cite{18} was applied to Fe-$3d$ states. The total energy $E$ in the LDA+$\textit{U}$ functional with the simplified formulation of Dudarev $et\ al.$ \cite{42} is written as

\begin{equation}
E=E_{LDA}+\frac{U}{2}\sum_{I,\sigma}[\boldsymbol{n}^{I,\sigma}(1-\boldsymbol{n}^{I,\sigma})]
\label{dftu},
\end{equation}

where $E_{LDA}$ is the LDA ground-state energy and $\boldsymbol{n}^{I,\sigma}$ is the occupation matrix of the atomic site $I$ with spin $\sigma$. The Hubbard parameter $U$ was computed using DFPT \cite{27} implemented in the Quantum ESPRESSO code \cite{43,44}. The convergence threshold for the response function is $1\times{10}^{-10}$ Ry. 

\begin{figure}[t]
\includegraphics[width=0.45\textwidth]{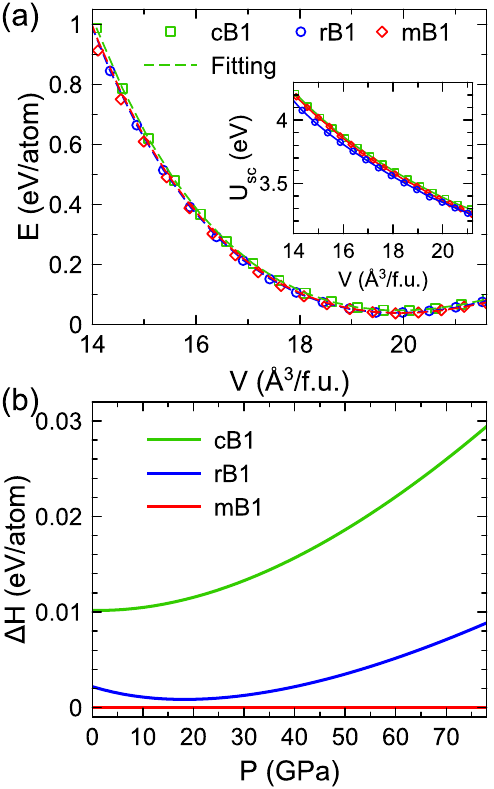}
\caption{\label{fig:fig2} (a) Relative energies of the three B1 phases. The dashed line gives results fitted with the third-order Birch-Murnaghan equation of state.  The inset shows the self-consistent Hubbard parameters as a function of the volume for the three phases. (b) Relative enthalpies of the three mB1 phases. These results correspond to static calculations.}
\end{figure}

Here, we automated an iterative scheme to obtain a self-consistent $\textit{U}_\text{sc}$ parameter while simultaneously optimizing the structure and desired spin state: Starting from an empirical $U$ of 4.3 eV, the energies of all possible occupation matrices for a spin state were computed. There are five possible occupation matrices corresponding to the HS state of ferrous iron with $3d^6$ configuration (S=2), while there are ten possibilities for the LS state (S=0). The electronic configuration, i.e., occupation matrix, with the lowest energy, was selected for further structural optimization of both lattice parameters and atomic positions. Then a new $U$ parameter is recalculated for further structural optimization. The process continued until mutual convergence of structure and U is achieved for a convergence threshold of 0.01 eV for the $U$ parameter and the convergence criteria mentioned above for structural optimizations.

\section{Results and Discussion}
\subsection{B1 phases}

Using the LDA+$\textit{U}_\text{sc}$ scheme, we first investigate FeO with the cB1, rB1, and mB1 structures. The lattices and antiferromagnetic configurations of rB1 and mB1 phases are shown in Fig.~\ref{fig:fig1}(a) and (b), respectively. The cubic structure is represented by the rhombohedral lattice with fixed lattice angles \cite{14}. The rB1 and mB1 structures were fully relaxed for the lattice parameters at each volume. After relaxation, the structures are symmetrized with a tight tolerance of 0.01 $\text{\AA}$ using the FINDSYM software \cite{47}, which confirms no symmetry change happens during the relaxation. cB1 lattice has no degree of freedom to be relaxed, so only self-consistent calculations are performed to obtain energy. Figure~\ref{fig:fig2}(a) shows the volume-dependent energy and self-consistent Hubbard parameters for three B1 phases. The obtained Hubbard parameters are very close among the three phases but show a strong dependence on the pressure. The energy-volume data are fitted by the third-order Birch-Murnaghan (BM) equation of state (EoS). The enthalpies are obtained from the static compression curves fitted to the BM-EoS and are shown in Fig.~\ref{fig:fig2}(b). At P=0 GPa, the mB1 structure is only 2.2 meV/atom (equivalent to $\sim$ 25K) lower than the rB1 structure. This is consistent with the fact that the mB1 phase was only observed at very low temperatures $T=10K$ \cite{31}. The mB1 phase has lower enthalpy than the cB1 and rB1 phases at all pressures in this static calculation. Vibrational effects may change this result at some pressure and temperature, but this effect was not investigated here. EoS parameters and bond lengths of the mB1 phase at the equilibrium volume $V_0$ is shown in Table~\ref{table:tab1}. The equilibrium volume and distorted bond lengths show very good agreement with experimental data \cite{31}. It’s even closer to experiments compared to the calculation using hybrid functional B3PW91 \cite{48}. Therefore the calculation with a self-consistent Hubbard parameter seems more predictive than the one with constant “magic number” 4.2 eV used in previous calculations \cite{48}. This result may change somewhat after the inclusion of vibrational effects, but it is clear the self-consistent scheme is necessary to obtain a Hubbard parameter for a predictive DFT+$\textit{U}$ calculation. 

\onecolumngrid

\clearpage

\begin{table}
\caption{\label{table:tab1} Comparison of structural information of the mB1 with previous results. $V_0$ is the equilibrium volume at $P=0\ GPa$. $K_0$ and $K_0^’$ are the bulk modulus and the derivative of bulk modulus, respectively. Calculations are all static.}
\centering
\begin{tabular} {  c | c | c | c | c | c  }
\hline
\hline
\textbf{ } Methods & V$_\text{0}$  & K$_\text{0}$  & K$_\text{0}^{'}$ & \multicolumn{2}{c}{Octahedron distortion} \\
\cline{5-6}
& (\AA$^\text{3}$/f.u.) & (GPa) & & Long bond (\AA) & Short bond (\AA) \\
\hline
Experiment Fe$_\text{0.99}$O at 10 K \cite{28} & 20.090 & - & - & 2.165 & 2.154 \\
\hline
LDA+$\textit{U}_{\text{sc}}$ & 19.915 & 188.1 & 4.01 & 2.171 & 2.144 \\
\hline
B3PW91 \cite{43} & 19.804 & 167.5 & 3.70 & 2.176 & 2.145 \\
\hline
\hline
\end{tabular}
\end{table}
\twocolumngrid

\begin{figure}[b]
\includegraphics[width=0.42\textwidth]{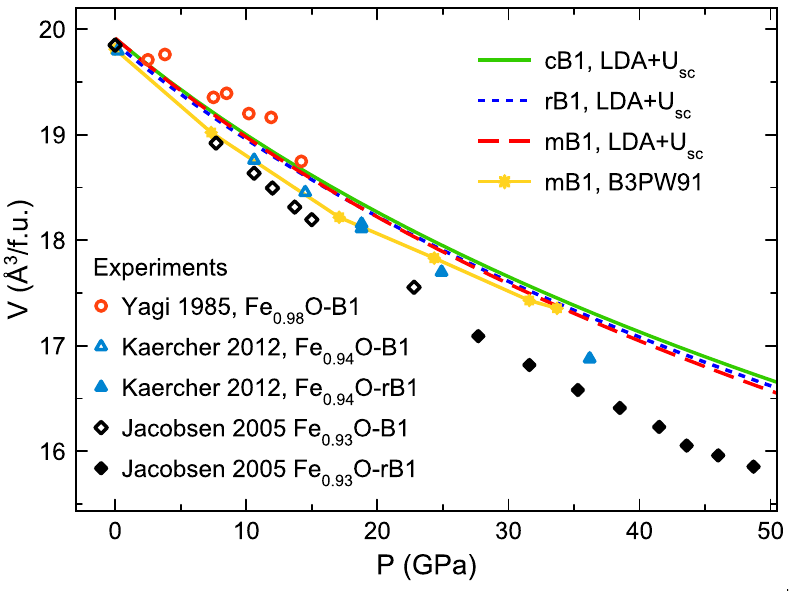}
\caption{\label{fig:fig3} Pressure-volume curves of B1 phases from calculations and experiments. The data of B3PW91 calculation is from Ref. \cite{48}. The experiments were performed at room temperature by Yagi $et\ al.$ \cite{1}, Jacobsen $et\ al.$ \cite{49} and Kaercher $et\ al.$ \cite{50}}
\end{figure}

Figure~\ref{fig:fig3} shows the compression curves of the B1 phases. The three polytypes (cB1, rB1 and mB1) show very similar compressive behavior, but it can be noticed that structures with more degrees of freedom ($dg$) to relax ($dg^{mB1} > dg^{rB1}> dg^{cB1}$) are more compressive, i.e., $\beta^{mB1} > \beta^{rB1}> \beta^{cB1}$ with $\beta$ being the compressibility. Comparing with results of a previous calculation using the hybrid functional B3PW91 \cite{48}, results are consistent except that the B3PW91 results deviated from ours in the range of 5-20 GPa. In fact, B3PW91 results $\textit{do not}$ fit well to the BM-EoS.

Next, we compare our B1 compression curve calculations with experimental data from a few different experiments \cite{1,49,50}, to the best of our knowledge. This comparison clearly shows that the compressive behavior of FeO depends quite strongly on this compound stoichiometry. Iron vacancies are common because Fe is a multivalent ion and frequently exists in the ferric form, (Fe$^{3+}$). The higher the iron vacancy concentration the higher the compressibility. The present calculations have no vacancies, therefore, they show smaller compressibility than the experimental data. Despite having no vacancies, our volumes are smaller than those reported for Fe$_\text{0.98}$O. This is caused by the static nature of our results. Inclusion of vibrational effects would likely improve agreement between theoretical and experimental data at 300 K \cite{51}. Considering these two effects, i.e., stoichiometry difference and vibrational effects, the current agreement between theoretical results and room temperature experimental data can be considered excellent. Therefore, LDA+$\textit{U}_\text{sc}$ can indeed well describe the structural properties of insulating B1 phases.

\begin{figure}[b]
\includegraphics[width=0.45\textwidth]{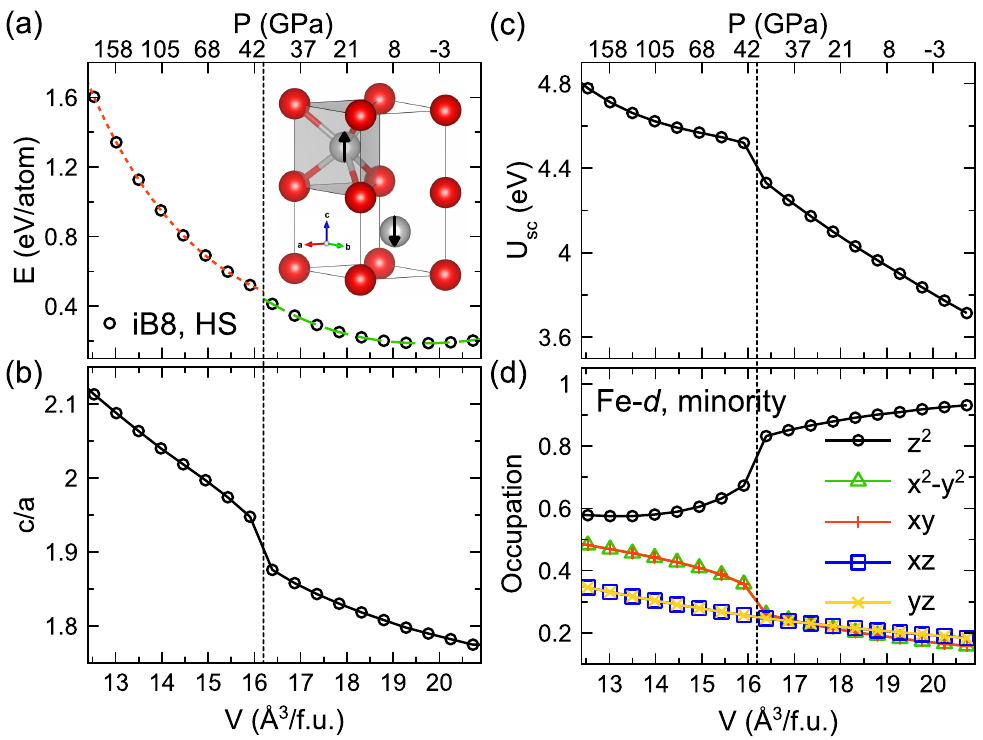}
\caption{\label{fig:fig4} (a) The energy-volume curves for iB8 in the HS state. The red dotted line and green dashed lines are the BM-EoS fitting for small volume and large volume results, respectively. The inset shows the atomic structure of the iB8-HS phase. Grey and red spheres are iron and oxygen, respectively. (b) The $c/a$ ratio,  (c) the self-consistent Hubbard parameters and (d) minority electron $d$-orbital occupancies vs. volume. The vertical lines indicate discontinuities. }
\end{figure}

\subsection{B8 phases}

\begin{figure}[t]
\includegraphics[width=0.4\textwidth]{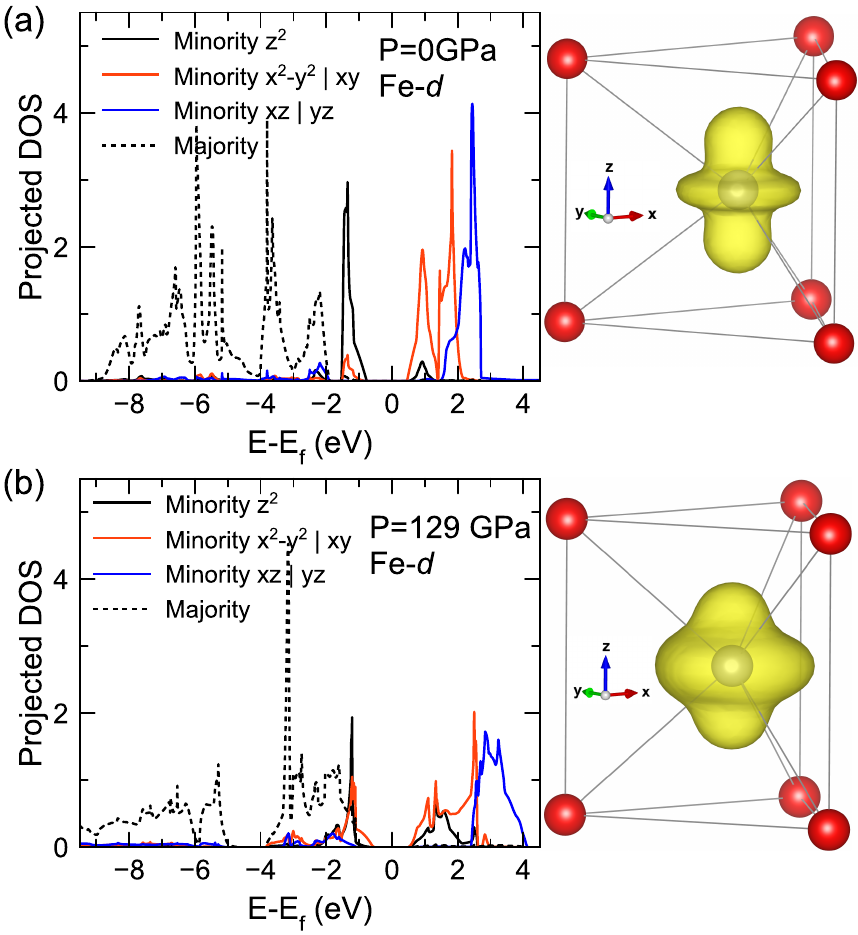}
\caption{\label{fig:fig5} Electronic density of state of Fe $d$-orbitals in the iB8-HS at (a) 0 GPa and (b) 129 GPa. The right panel shows the occupied Fe-$d$ minority orbital. The iso-surface threshold values are the same in (a) and (b). }
\end{figure}

Figure~\ref{fig:fig4} shows the LDA+$\textit{U}_\text{sc}$ results for the iB8 HS state (S=2). The antiferromagnetic configuration consists of alternating up and down spins along the $c$ direction. The energy-volume curve in Fig.~\ref{fig:fig4}(a) shows an unexpected discontinuity at 16.2 $\text{\AA}^3/\text{f.u}$. near 40 GPa. It requires two BM-EoS fittings, one for the high-pressure and one for the low-pressure results. The $c/a$ ratio and the computed self-consistent Hubbard parameter also show a consistent discontinuity at the same volume (pressure) in Fig.~\ref{fig:fig4}(b) and (c). The origin of this behavior can be tracked to the orbital occupancy of the minority d electron in Fe$^{2+}$ in the HS state. In the iB8 HS structure the majority spin up bands are completely full so only one electron enters in the spin-down bands ($d_\uparrow^5d_\downarrow^1$ electronic configuration). The projected atomic orbital occupancies for the minority electron are plotted vs. volume in Fig.~\ref{fig:fig4}(d). At large volumes, the occupied state is mainly a $z^2$–type orbital (short for $3z^2-r^2$), but the $z^2$ orbital occupancy decreases under compression. Note that the $z$-axis is aligned along the crystal $c$ axis. For $V\le 16.2 \text{\AA}^3/\text{f.u.}$ ($\sim$ 40 GPa), the occupied $d$-band shows a strong mixed $z^2/x^2-y^2/xy$ character. To clarify this situation, we show in Fig.~\ref{fig:fig5} the projected electronic density of state for Fe $d$-orbitals in the iB8-HS phase at low and high pressures. One can see a greater orbital mixing at higher pressures by comparing Figs.~\ref{fig:fig5}(a) and (b). The shapes of the occupied localized orbitals are also shown in Fig.~\ref{fig:fig5}, indicating that the minority electronic density shifts to the $x-y$ plane, and the $c$-axis compressibility increases after that. The bandgap of $\sim$ 1.0 eV is essentially pressure independent.

This electronic transition in the iB8-HS phase has structural effects, but it is a metastable transition, i.e., the stable phase at $\sim$ 40 GPa is the mB1 phase. Nevertheless, this type of electronic transition in iron is not uncommon. For example, Mössbauer spectroscopy detected the increased presence of a second type of ferrous iron with higher quadrupole splitting (3.5 mm/sec) in (Mg,Fe)SiO$_3$-perovskite, i.e., bridgmanite, starting at 30 GPa and ending at 60 GPa \cite{52}. Such change was attributed to an HS to intermediate spin (IS) state (S=1) change, which happens continuously with pressure. It was later shown that such state change consists in the change of orbital occupancy of the minority d-electron\cite{53}, which is also accompanied by a change in compressibility of bridgmanite\cite{54}. The total spin, S=2, was not altered throughout this electronic transition\cite{55}. 

\begin{figure}
\includegraphics[width=0.85\textwidth]{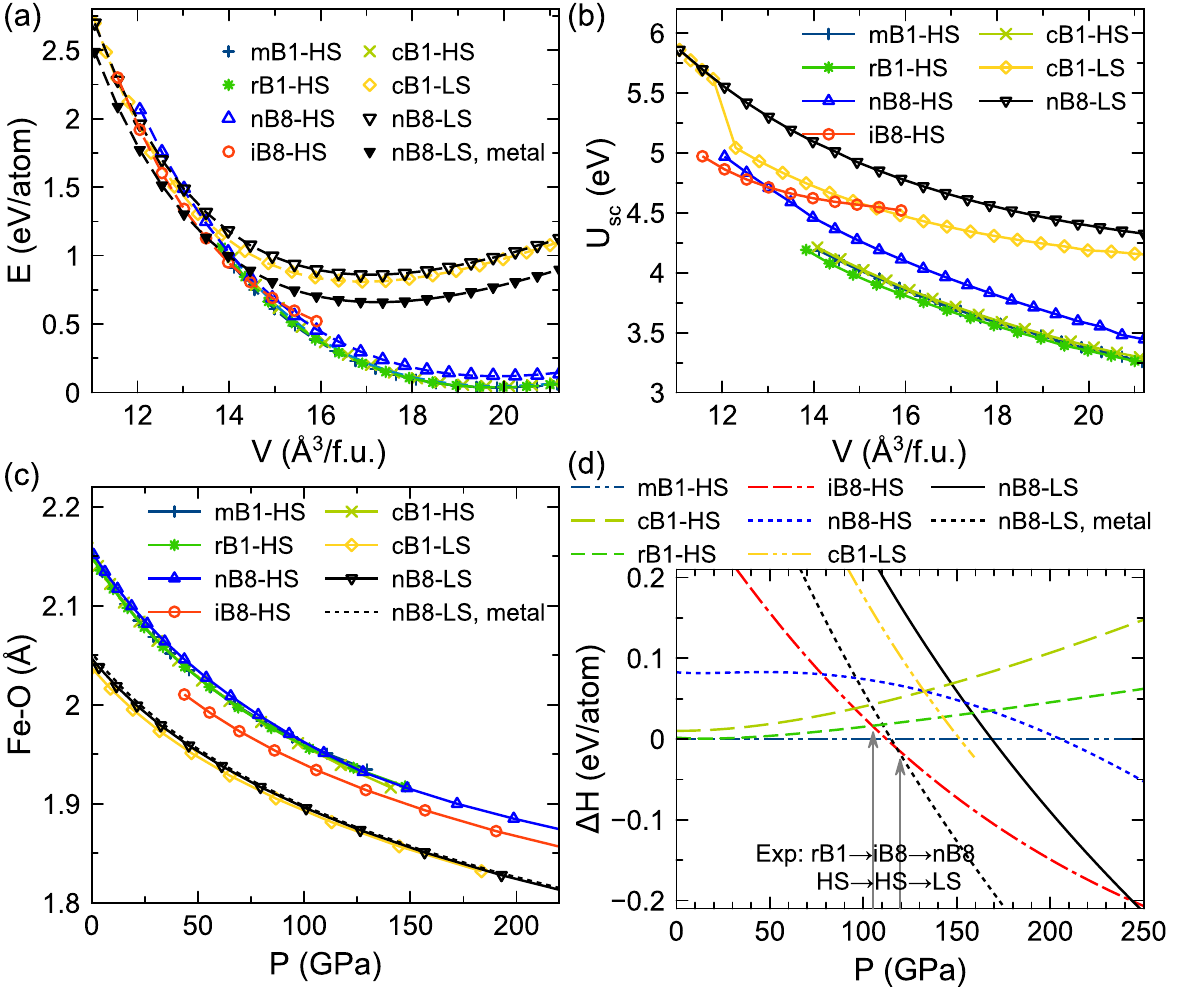}
\caption{\label{fig:fig6} (a) Energy-volume relation for relevant B1 and B8 phases. The curves of mB1-HS, cB1-HS and rB1-HS are almost overlapped on this scale. (b) Self-consistent Hubbard parameter vs. volume for the same phases. (c) Average Fe-O bond lengths vs. pressure. (d) Relative enthalpies using the data of mB1-HS phase as the reference. Arrows indicate experimental transition pressures of rB1-HS to iB8-HS to nB8-LS. The “nB8-LS, metal” indicates the one using Mermin functional with $T_{el} =7,000 K$. }
\end{figure}

To compare the relative stability of B1 and B8 phases, LDA+$\textit{U}_\text{sc}$ calculations were performed for nB8, iB8, and B1 phases in both HS and LS states. Because the ferromagnetic configuration shows higher energy than the antiferromagnetic configurations systematically at all volumes, only antiferromagnetic configurations are reported here. Because the iB8-LS phase had high energy and is mechanically unstable, consistent with a previous finding \cite{26}, the iB8-LS phase is disregarded. By checking the symmetry of the relaxed structures, we find the rB1-LS and mB1-LS phases spontaneously transform to the cB1-LS structure, so there are no rB1-LS or mB1-LS states. More importantly, these B1-LS phases transform to the nB8-LS when volumes are smaller than 12.0 $\text{\AA}^3$/f.u. (i.e. $\sim$ 160 GPa). All the LDA+$\textit{U}_\text{sc}$ results for B1 and B8 phases are shown in Fig.~\ref{fig:fig6}.

The self-consistent Hubbard parameters are shown in Fig.~\ref{fig:fig6}(b) and display strong dependences on pressure, structure, and spin states. LS states systematically have larger Hubbard parameters than HS states, regardless of structure. Figure~\ref{fig:fig6}(c) shows average Fe-O bond lengths versus pressure. They are almost indistinguishable in mB1-HS, cB1-HS, rB1-HS, and nB8-HS phases and in the cB1-LS and nB8-LS phases. Iron in these phases is all octahedrally coordinated (see Fig.~\ref{fig:fig1}). The pressure dependence of the average FeO bond lengths in iB8-HS is different from those of other phases. In this phase, the iron coordination polyhedron is a triangular prism.  Therefore, phases with iron of the same coordination polyhedron and spin states have almost indistinguishable Fe-O bond-lengths at the same pressure.

\begin{figure}
\includegraphics[width=0.40\textwidth]{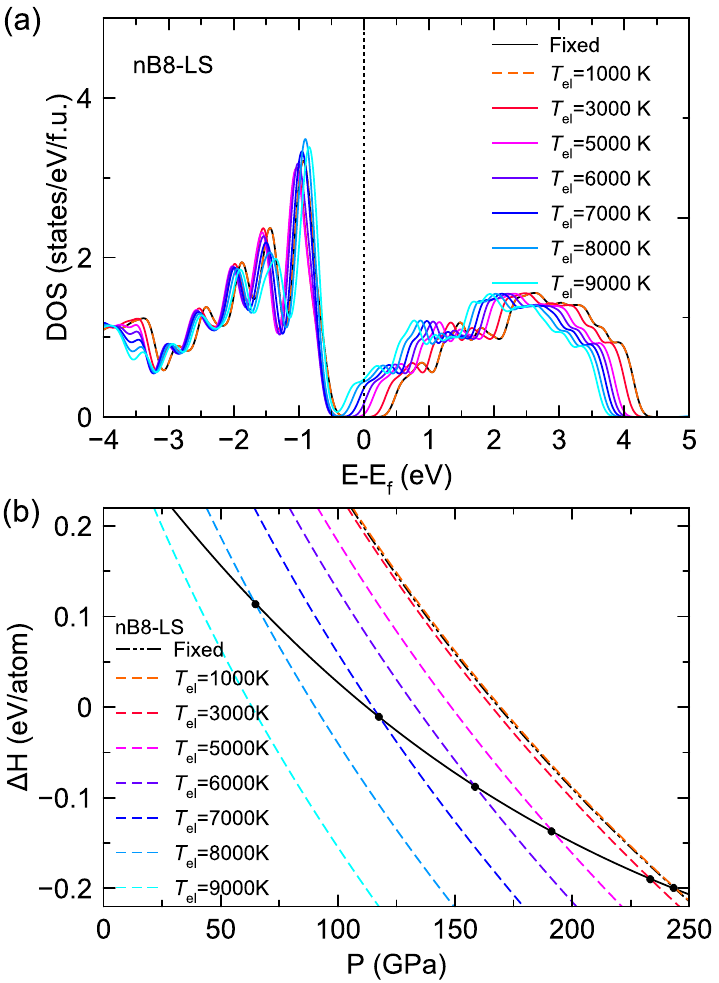}
\caption{\label{fig:fig7} (a) The density of state of nB8-LS at 110 GPa using different occupation schemes. “Fixed” corresponds to the fixed-occupation scheme for the insulating state. The T$_{el}$ corresponds to the broadening in the Fermi-Dirac smearing. (b) Relative enthalpies of iB8-HS and nB8-LS with smearing. The enthalpy of mB1-HS is used as the reference. The dots indicate the transition pressure. }
\end{figure}

The relative stability of these phases is shown in Fig.~\ref{fig:fig6}(d). The mB1 phase is the most stable up to 110 GPa at T=0K. Because only the rB1 phase (stable at room temperature) was reported in the previous high-pressure experiments, we focus on the phase transitions from the latter. The current LDA+$\textit{U}_\text{sc}$ calculations produce two phase transitions: rB1-HS to iB8-HS at 105 GPa and iB8-HS to nB8-LS at 245 GPa, which are all insulating phases. The first transition from rB1-HS to iB8-HS is very consistent with the experimental transition pressure of $\sim$105 GPa at room temperature \cite{2,30,32}. However, the second transition from iB8-HS to nB8-LS happens at a significantly larger pressure than the experimental transition pressure of 120 GPa\cite{39}. By examining the electronic density of state shown in Fig.~\ref{fig:fig7}(a), we find this nB8-LS state to be insulating with a bandgap of $\sim$ 0.8 eV, which contrasts with the metallic nB8-LS phase obtained experimentally. 

\onecolumngrid

\clearpage

\twocolumngrid

The current LDA+$\textit{U}_\text{sc}$ scheme essentially promotes the insulating state by penalizing the metallic state with increasing $\textit{U}_\text{sc}$ at higher pressures. To evade this problem, after obtaining $\textit{U}_\text{sc}$, we continue calculations on this phase using the Mermin functional\cite{56,57}, with orbitals occupancies given by the Fermi-Dirac distribution and electronic entropic contribution included in the total “free energy” calculation. We test the outcome of this strategy for variable “electronic temperature,” T$_{el}$. As shown in Fig.~\ref{fig:fig7}, by increasing T$_{el}$, the Fermi level gradually shifts to the conduction band, and the transition pressure from iB8-HS to nB8-LS decreases. With T$_{el}$ ~ 7,000 K, the nB8-LS phase becomes a metal. The iB8-HS to nB8-LS transition pressure is also lowered to $\sim$120 GPa, in agreement with the experimental value of 120 GPa. The metallization of the nB8-LS phase and the iB8-HS to nB8-LS transition pressure can be controlled by the magnitude of T$_{el}$ and the value of the Hubbard parameter, though freely manipulating the latter is not under consideration here. This artificially large T$_{el}$ and its co-dependence on the Hubbard $U$ value points to the necessity to render the DFT+$\textit{U}_\text{sc}$ scheme more flexible to improve its description of metallic ground states.

\begin{figure}
\includegraphics[width=0.4\textwidth]{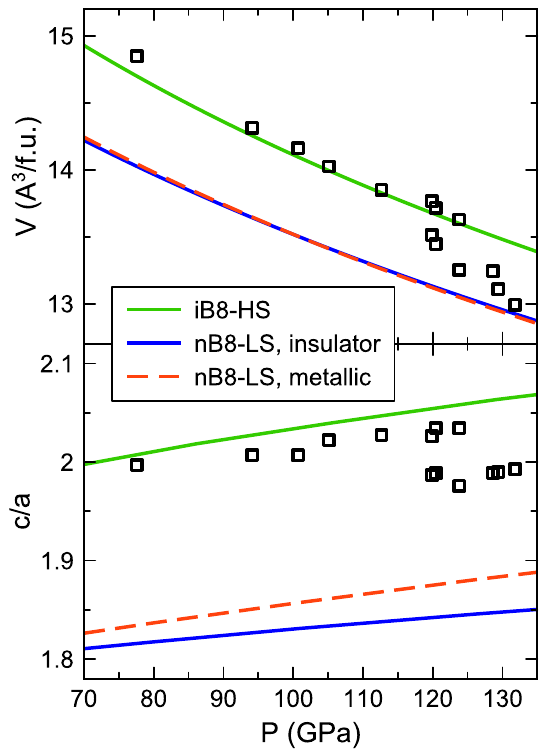}
\caption{\label{fig:fig8} The volume and $c/a$ ratio as a function of pressure for iB8-HS and nB8-LS phases. The experimental data includes both iB8 and nB8 data with a transition pressure at 120 GPa [39]. }
\end{figure}

The structures of iB8-HS and nB8-LS are also compared with experimental data in Fig.~\ref{fig:fig8}. Calculated volume and $c/a$ ratio for the iB8-HS state agree well with the experimental data obtained at room temperature\cite{39}. Because the iB8-HS was found experimentally to be insulating, the current LDA+$\textit{U}_\text{sc}$ scheme indeed describes its electronic and structural characters quite well. However, the insulating nB8-LS phase shows a significant deviation from experimental values in the $c/a$ ratio. Treating the system as metallic by using the Mermin functional with an electronic temperature T$_{el}$, the $c/a$ ratio gets closer to the experimental values.

\clearpage

The current calculation does not include spin-orbit coupling (SOC) effect because it is believed to be insignificant in the FeO system \cite{13}. To confirm this idea, we examine the SOC effect on the transition pressure from rB1-HS to iB8-HS in the Supplemental Material. We find that SOC increases the total energy of both rB1-HS and iB8-HS phases by $\sim$ 0.04 eV/atom almost uniformly in the volume range explored. Therefore, it does not change the transition pressure noticeably.

\section{Conclusion}
In summary, we have shown that with a careful determination of the Hubbard parameter, LDA+$\textit{U}_\text{sc}$ can well describe phase relations among and compression curves of $insulating$ FeO phases. The self-consistent Hubbard parameters $\textit{U}_\text{sc}$ depends strongly on pressure, structure, and spin state, varying typically by 1-2 eV, as shown in Fig.~\ref{fig:fig6}. Therefore, LDA+$\textit{U}$ calculations with constant or artificially tuned $U$ values may not be able to capture structural and electronic properties fully. For B1 phases, current calculations confirm the mB1 phase should be the ground state at T=0 K, while the rB1 phase shows similar enthalpy to that of the mB1 phase. Therefore, the mB1 phase could only be observed experimentally at low-temperatures\cite{31}. We also found a metastable electronic transition in the iB8-HS phase at $\sim$ 40 GPa. This transition consists of a change in orbital occupancy of the minority $d$-electron, a phenomenon similar to that observed in ferrous iron in (Mg,Fe)SiO$_3$ between 30 GPa and 60 GPa\cite{52,53}. The zero-temperature phase boundary between rB1-HS and iB8-HS is 105 GPa, which is very consistent with the room-temperature observation at 105 GPa\cite{30,32}. The equilibrium volume of mB1-HS, the compression curve of B1-HS and iB8-HS agrees very well with experimental measurements. The only discrepancy between current calculations and experiments is on the nB8-LS state that overestimates the transition pressure and underestimates the structural $c/a$ ratios compared to experiments. This is because LDA+$\textit{U}_\text{sc}$ produces an insulating state for nB8 while it should be metallic. By using the Mermin functional $a\ posteriori$, i.e., without further changing $U$, and an artificially large electronic temperature T$_{el}$ $\sim$ 7,000 K, the LDA+$\textit{U}$ calculation produces a metallic state, improves structural properties, and produces an iB8-HS to nB8-LS transition pressure in good agreement with experiments. All these properties can also be modified by changing the Hubbard parameter. This artificially large T$_{el}$ and its co-dependence on the Hubbard $U$ value points to the type of modification necessary in the DFT+$\textit{U}_\text{sc}$ scheme to address the metallic state that includes electronic entropy contributions in the electronic free energy.

\section{acknowledgments}
The authors thank Y. Yao, V. Antropov and K.-M. Ho for helpful discussion. This work was funded in part by National Science Foundation award EAR-1918126 (Y. S.) and in part by the US Department of Energy award DESC0019759 (R.M.W.). This work used the Extreme Science and Engineering Discovery Environment (XSEDE), USA, which was supported by the National Science Foundation, USA Grant Number ACI-1548562. Computations were performed on Stampede2, the flagship supercomputer at the Texas Advanced Computing Center (TACC), The University of Texas at Austin generously funded by the National Science Foundation (NSF) through award ACI-1134872. 

\twocolumngrid

\renewcommand{\bibnumfmt}[1]{[#1]}
\bibliographystyle{apsrev4-1}
%


\pagebreak
\widetext
\begin{center}
\textbf{\large Supplemental Material for ``LDA+$\textit{U}_\text{sc}$ calculations of phase relations in FeO''} 
\end{center}

\setcounter{equation}{0}
\setcounter{figure}{0}
\setcounter{table}{0}
\makeatletter
\renewcommand{\theequation}{S\arabic{equation}}
\renewcommand{\thefigure}{S\arabic{figure}}
\renewcommand{\bibnumfmt}[1]{[S#1]}
\renewcommand{\citenumfont}[1]{S#1}
\renewcommand{\thesection}{S\arabic{section}}
\twocolumngrid

\begin{figure}[h]
\includegraphics[width=0.48\textwidth]{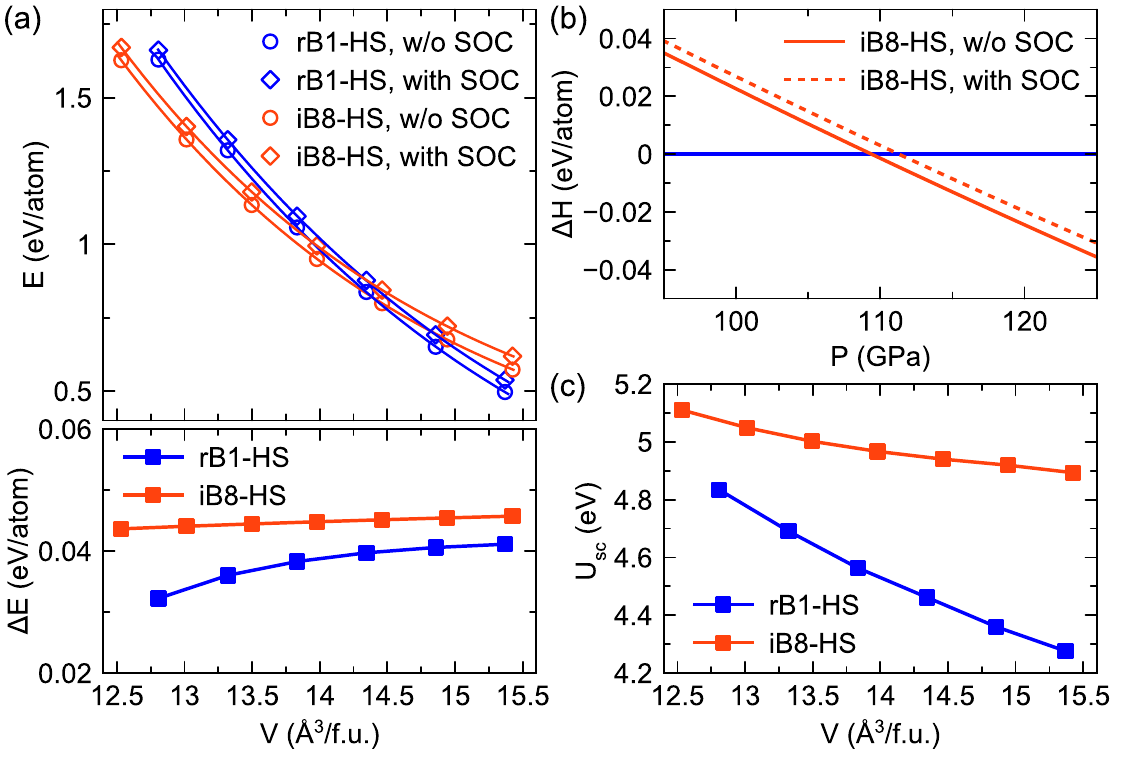}
\caption{\label{fig:figS1} (a) The upper panel shows the Energy-volume relation for rB1-HS and iB8-HS phases with and without spin-orbital coupling. The solid lines are the fitting results with the third-order Birch-Murnaghan equation of state. The lower panel shows the energy difference caused by the inclusion of SOC, i.e. $\Delta E=E_{SOC}-E$. (b) Relative enthalpies using the data of rB1-HS phase as the reference. (c) The self-consistent $U_{sc}$ value. }
\end{figure}

Here, we examine the impact of the spin-orbital coupling (SOC) on the rB1-HS to iB8-HS transition pressure. The SOC calculations are performed with a rotationally invariant form of LSDA+$U$+$J$ functional \cite{s1}. The parameter $J$ is fixed at $J_c=0.9$ and $U=U_{sc}+ J_c$, where $U_{sc}$ is obtained from self-consistent calculations with the simplified formulation of Dudarev \textit{et al.} \cite{s2}, as described in the main text. The fully relativistic pseudopotential is employed for the SOC calculations.

Figure \ref{fig:figS1}(a) shows that the inclusion of SOC systematically increases the energy $\sim$ 0.04 eV/atom, with a small dependence on pressure and phase. Because the SOC effect is similar on both  rB1-HS and iB8-HS phases, it  only changes the transition pressure $\sim$ 2GPa of $\sim$ 110GPa, as shown in Fig. \ref{fig:figS1}(b). Therefore, while SOC can indeed affect the results of total energy calculation, the impact on transition pressure is small.

Due to the change of pseudopotentials the self-consistent $U_{\text{sc}}$ is recalculated for both phases. The new values are provided in Fig. \ref{fig:figS1}(c). The current SOC calculation is not fully self consistent because of the presumed $J$ value. A more sophisticated calculation may be carried out using a recently developed noncollinear LSDA+$U$ technique \cite{s3}.

\bibliographystyle{apsrev4-1}

\begin{thebibliography}{52}%
\bibitem{1}	T. Yagi, T. Suzuki, and S. I. Akimoto, J. Geophys. Res. 90, 8784 (1985).
\bibitem{2}	Y. Fei and H. K. Mao, Science 266, 1678 (1994).
\bibitem{3}	R. E. Cohen, I. I. Mazin, and D. G. Isaak, Science 275, 654 (1997).
\bibitem{4}	B. Lavina, P. Dera, E. Kim, Y. Meng, R. T. Downs, P. F. Weck, S. R. Sutton, and Y. Zhao, Proc. Natl. Acad. Sci. 108, 17281 (2011).
\bibitem{5}	B. Lavina and Y. Meng, Sci. Adv. 1, e1400260 (2015).
\bibitem{6}	Q. Hu, D. Y. Kim, W. Yang, L. Yang, Y. Meng, L. Zhang, and H.-K. Mao, Nature 534, 241 (2016).
\bibitem{7}	K. Terakura, T. Oguchi, A. R. Williams, and J. Kübler, Phys. Rev. B 30, 4734 (1984).
\bibitem{8}	A. R. Oganov, Y. Ma, A. O. Lyakhov, M. Valle, and C. Gatti, Rev. Mineral. Geochemistry 71, 271 (2010).
\bibitem{9}	S. Q. Wu, M. Ji, C. Z. Wang, M. C. Nguyen, X. Zhao, K. Umemoto, R. M. Wentzcovitch, and K. M. Ho, J. Phys. Condens. Matter 26, 035402 (2014).
\bibitem{10}	A. R. Oganov, Y. Ma, C. W. Glass, and M. Valle, Psi-k Newsl. 84, 142 (2007).
\bibitem{11}	G. L. Weerasinghe, C. J. Pickard, and R. J. Needs, J. Phys. Condens. Matter 27, 455501 (2015).
\bibitem{12}	A. R. Oganov, C. J. Pickard, Q. Zhu, and R. J. Needs, Nat. Rev. Mater. 4, 331 (2019).
\bibitem{13}	I. Mazin and V. Anisimov, Phys. Rev. B 55, 12822 (1997).
\bibitem{14}	M. Cococcioni and S. de Gironcoli, Phys. Rev. B 71, 035105 (2005).
\bibitem{15}	A. Georges, G. Kotliar, W. Krauth, and M. J. Rozenberg, Rev. Mod. Phys. 68, 13 (1996).
\bibitem{16}	A. O. Shorikov, Z. V. Pchelkina, V. I. Anisimov, S. L. Skornyakov, and M. A. Korotin, Phys. Rev. B 82, 195101 (2010).
\bibitem{17}	N. Lanatà, T.-H. Lee, Y.-X. Yao, V. Stevanović, and V. Dobrosavljević, Npj Comput. Mater. 5, 30 (2019).
\bibitem{18}	V. I. Anisimov, J. Zaanen, and O. K. Andersen, Phys. Rev. B 44, 943 (1991).
\bibitem{19}	H. J. Kulik, M. Cococcioni, D. A. Scherlis, and N. Marzari, Phys. Rev. Lett. 97, 103001 (2006).
\bibitem{20}	H. Hsu, K. Umemoto, Z. Wu, and R. M. Wentzcovitch, Rev. Mineral. Geochemistry 71, 169 (2010).
\bibitem{21}	T. Tsuchiya, R. M. Wentzcovitch, C. R. S. da Silva, and S. de Gironcoli, Phys. Rev. Lett. 96, 198501 (2006).
\bibitem{22}	R. M. Wentzcovitch, J. F. Justo, Z. Wu, C. R. S. da Silva, D. A. Yuen, and D. Kohlstedt, Proc. Natl. Acad. Sci. 106, 8447 (2009).
\bibitem{23}	M. Cococcioni and N. Marzari, Phys. Rev. Mater. 3, 033801 (2019).
\bibitem{24}	A. Floris, I. Timrov, B. Himmetoglu, N. Marzari, S. De Gironcoli, and M. Cococcioni, Phys. Rev. B 101, 064305 (2020).
\bibitem{25}	S. A. Gramsch, R. E. Cohen, and S. Y. Savrasov, Am. Mineral. 88, 257 (2003).
\bibitem{26}	K. Persson, A. Bengtson, G. Ceder, and D. Morgan, Geophys. Res. Lett. 33, L16306 (2006).
\bibitem{27}	I. Timrov, N. Marzari, and M. Cococcioni, Phys. Rev. B 98, 085127 (2018).
\bibitem{28}	M. Murakami, K. Hirose, S. Ono, T. Tsuchiya, M. Isshiki, and T. Watanuki, Phys. Earth Planet. Inter. 146, 273 (2004).
\bibitem{29}	K. Ohta, R. E. Cohen, K. Hirose, K. Haule, K. Shimizu, and Y. Ohishi, Phys. Rev. Lett. 108, 026403 (2012).
\bibitem{30}	T. Irifune and T. Tsuchiya, Phase Transitions and Mineralogy of the Lower Mantle (Elsevier B.V., 2015).
\bibitem{31}	H. Fjellvag, B. C. Hauback, T. Vogt, and S. Stolen, Am. Mineral. 87, 347 (2002).
\bibitem{32}	H. Ozawa, K. Hirose, S. Tateno, N. Sata, and Y. Ohishi, Phys. Earth Planet. Inter. 179, 157 (2010).
\bibitem{33}	I. I. Mazin, Yingwei Fei, R. Downs, and R. Cohen, Am. Mineral. 83, 451 (1998).
\bibitem{34}	Z. Fang, K. Terakura, H. Sawada, T. Miyazaki, and I. Solovyev, Phys. Rev. Lett. 81, 1027 (1998).
\bibitem{35}	M. P. Pasternak, R. D. Taylor, R. Jeanloz, X. Li, J. H. Nguyen, and C. A. Mc Cammon, Phys. Rev. Lett. 79, 5046 (1997).
\bibitem{36}	J. Badro, V. V. Struzhkin, J. Shu, R. J. Hemley, H. K. Mao, C. C. Kao, J. P. Rueff, and G. Shen, Phys. Rev. Lett. 83, 4101 (1999).
\bibitem{37}	A. Mattila, J. P. Rueff, J. Badro, G. Vankó, and A. Shukla, Phys. Rev. Lett. 98, 98 (2007).
\bibitem{38}	K. Ohta, K. Hirose, K. Shimizu, and Y. Ohishi, Phys. Rev. B 82, 174120 (2010).
\bibitem{39}	H. Ozawa, K. Hirose, K. Ohta, H. Ishii, N. Hiraoka, Y. Ohishi, and Y. Seto, Phys. Rev. B 84, 134417 (2011).
\bibitem{40}	D. M. Sherman and H. J. F. Jansen, Geophys. Res. Lett. 22, 1001 (1995).
\bibitem{41}	I. Leonov, Phys. Rev. B 92, 085142 (2015).
\bibitem{42}	S. Dudarev and G. Botton, Phys. Rev. B 57, 1505 (1998).
\bibitem{43}	P. Giannozzi, S. Baroni, N. Bonini, M. Calandra, R. Car, C. Cavazzoni, D. Ceresoli, G. L. Chiarotti, M. Cococcioni, I. Dabo, A. Dal Corso, S. de Gironcoli, S. Fabris, G. Fratesi, R. Gebauer, U. Gerstmann, C. Gougoussis, A. Kokalj, M. Lazzeri, L. Martin-Samos, N. Marzari, F. Mauri, R. Mazzarello, S. Paolini, A. Pasquarello, L. Paulatto, C. Sbraccia, S. Scandolo, G. Sclauzero, A. P. Seitsonen, A. Smogunov, P. Umari, and R. M. Wentzcovitch, J. Phys. Condens. Matter 21, 395502 (2009).
\bibitem{44}	P. Giannozzi, O. Andreussi, T. Brumme, O. Bunau, M. B. Nardelli, M. Calandra, R. Car, C. Cavazzoni, D. Ceresoli, M. Cococcioni, and others, J. Phys. Condens. Matter 29, 465901 (2017).
\bibitem{45}	D. Vanderbilt, Phys. Rev. B 41, 7892 (1990).
\bibitem{46}	K. Umemoto, R. M. Wentzcovitch, Y. G. Yu, and R. Requist, Earth Planet. Sci. Lett. 276, 198 (2008).
\bibitem{47}	H. T. Stokes and D. M. Hatch, J. Appl. Cryst. 38, 237 (2005).
\bibitem{48}	T. Eom, H. K. Lim, W. A. Goddard, and H. Kim, J. Phys. Chem. C 119, 556 (2015).
\bibitem{49}	S. D. Jacobsen, J. F. Lin, R. J. Angel, G. Shen, V. B. Prakapenka, P. Dera, H. K. Mao, and R. J. Hemley, J. Synchrotron Radiat. 12, 577 (2005).
\bibitem{50}	P. Kaercher, S. Speziale, L. Miyagi, W. Kanitpanyacharoen, and H. R. Wenk, Phys. Chem. Miner. 39, 613 (2012).
\bibitem{51}	R. M. Wentzcovitch, Y. G. Yu, and Z. Wu, Rev. Mineral. Geochemistry 71, 59 (2010).
\bibitem{52}	C. McCammon, I. Kantor, O. Narygina, J. Rouquette, U. Ponkratz, I. Sergueev, M. Mezouar, V. Prakapenka, and L. Dubrovinsky, Nat. Geosci. 1, 684 (2008).
\bibitem{53}	H. Hsu, K. Umemoto, P. Blaha, and R. M. Wentzcovitch, Earth Planet. Sci. Lett. 294, 19 (2010).
\bibitem{54}	G. Shukla, Z. Wu, H. Hsu, A. Floris, M. Cococcioni, and R. M. Wentzcovitch, Geophys. Res. Lett. 42, 1741 (2015).
\bibitem{55}	H. Hsu and R. M. Wentzcovitch, Phys. Rev. B 90, 195205 (2014).
\bibitem{56}	N. D. Mermin, Phys. Rev. 137, A1441 (1965).
\bibitem{57}	R. M. Wentzcovitch, J. L. Martins, and P. B. Allen, Phys. Rev. B 45, 11372 (1992).


\end{thebibliography}

\begin{thebibliography}{3}%

\bibitem{s1}	A. I. Liechtenstein, V. I. Anisimov, and J. Zaanen, Phys. Rev. B 52, R5467 (1995).
\bibitem{s2}	S. L. Dudarev, G. A. Botton, S. Y. Savrasov, C. J. Humphreys, and A. P. Sutton, Phys. Rev. B 57, 1505 (1998).
\bibitem{s3}	S. L. Dudarev, P. Liu, D. A. Andersson, C. R. Stanek, T. Ozaki, and C. Franchini, Phys. Rev. Mater. 3, 083802 (2019).
\end{thebibliography}

\end{document}